# Liquid Metal Molecular Scissors

**Liangfei Duan,[†a, b] Tong Zhou [†a], Huiqin Yang, [b] Weihua Mu,[b] Zhongshan Deng,[c] Jing Liu,[*c,d] Qingju Liu,[*a]**

[a] Yunnan Key Laboratory for Micro/Nano Materials & Technology, International Joint Research Center for Optoelectronic and Energy Materials, School of Materials and Energy, Yunnan University, Kunming 650091, China.

[b] School of Faculty of Chemistry and Chemical Engineering, Yunnan Normal University, Kunming 650500, China.

[c] CAS Key Laboratory of Cryogenics and Beijing Key Laboratory of Cryo- Biomedical Engineering, Technical Institute of Physics and Chemistry, Chinese Academy of Sciences, Beijing 100190, China.

[d] Department of Biomedical Engineering, School of Medicine, Tsinghua University, Beijing 100084, China

[†] **Co-First authors:**

***Corresponding authors:**

Jing Liu, E-mail, jliu@mail.ipc.ac.cn

Qingju Liu E-mail, qjliu@ynu.edu.cn

## Abstract

Molecules are the smallest unit in matters that can exist independently, relatively stable, and maintain physical and chemical activities. The atomic species, alignment commands, and chemical bonds are key factors to dominate their structures and properties. Here we disclosed a general chemistry effect that the liquid metals can directly cut off oxygen-containing groups in various molecular matters at room temperature, and then recombine the remaining groups to form functional materials including nano semiconductors. Based on this unique mechanism, we proposed a basic tool and named it as liquid metal scissors for molecular directional clipping and functional transformation. As proof-of-concept, we demonstrated the capabilities of eGaIn scissors made of Ga and In particles, and revealed that the Ga on the surface of eGaIn could directly snatch oxygen atoms from various targeted substances such as $H_2O$, $CO_2$ or $CH_3OH$ molecules to form gallium oxides. As illustration, after clipping, the remaining hydrogen atoms of $H_2O$ molecules recombined to form $H_2$, while the remaining groups of $CH_3OH$ lead to $H_2$, carbon quantum dots, and other related substances. If needed, more molecules can also be manipulated via such scissors. This finding refreshes the basic knowledge of chemistry and suggests easygoing ways for molecular weaving, which may break up the limitations and single features of molecular substances. It also opens up a universal route for innovating future molecular chemical engineering, life science, energy and environment, and biomedicine.

## INTRODUCTION

The basic constituent unit of matters mainly include atoms and molecules, and the molecules are the smallest unit in matters that can exist independently, relatively stable, and maintain physical and chemical properties. [1] The molecules are composed of atoms, while the atoms combine into a variety of molecules under certain force, alignment commands, and chemical bonds. [2] Molecular structures strongly determine the performance and application of substances, including the alignment order of atoms and chemical bond types. [3] Molecular cutting, editing and recombination are an effective strategy to construct emerging functional substances, which provides infinite possibilities for cutting-edge sciences such as biopolymers, small molecular substances, life sciences, advanced materials, biomedicine and energy environment through directional chemical editing means. [4] Historically, the ability to control and manipulate matters at molecular levels has significantly promoted human life and science advancement. [5] However, as it was pointed out, [6] current molecular editing and assembly generally rely heavily on cumbersome processes, complex reagents, and harsh reaction conditions, which turn out to be a large obstacle.

Among the many exciting functional matters, liquid metals (LMs) are emerging as new generation materials with rather unique physical and chemical behaviors. [7] LMs are safe and non-toxic, have high boiling points, reflectivities, good thermal and electrical conductivities, high flexibility, fluidity, self-healing capability and remain in liquid state at room temperature. [8] Particularly, the surface features of LMs significantly differ from solid metals, and the oxide shells with a thickness of ≈ 1-3 nm are formed on their surface spontaneously and instantaneously upon exposure to atmospheric oxygen. [9] The surface of liquid metals is sensitive to foreign matters. [10] Based on these tactics, it is speculated that LMs contain plenty of outstanding surface behaviors, and their functional capabilities and increasing application scopes would therefore be endowed through surface science discovery and practical exploration. [11] Unfortunately, investigation on the surface issues of LMs is still in the infancy and there exist big gaps within the knowledge of related mechanisms. [12]

In this article, based on a generalized finding that liquid metals can quickly achieve directional cutting, editing, and recombination of molecular substances via oxygen capture mechanism, we proposed to construct a basic tool of liquid metal molecular scissors. It can be made of gallium (Ga), its alloy with indium (In), or more other suitable low melting point alloys. In this sense, a particular tool can be called, say gallium molecular scissors. As disclosed in our experiments, the room temperature liquid metals such as gallium or its alloy have excellent surface activity, which can directionally cut off specific functional groups in molecules, and break the original stability of the chemical bonds so as to achieve the purpose of molecular weaving. As a proof-of-concept and typical example, the Ga and In particles are mixed together to form eGaIn liquid metal scissors. Without loosing any generality and also for brevity, we choose to illustrate several representative molecular substances although more candidates can also be tested. As our experiments disclosed, the excellent surface activity of eGaIn make them react with the inorganic molecule of $H_2O$ and organic molecule of $CH_3OH$ spontaneously and instantaneously. And more molecular substances such as $CO_2$ can also be manipulated via this way. The relevant mechanisms and potential applications have been elaborated systematically. This finding opens insightful and promising route for molding future molecule chemical engineering, life science, energy and environment, and biomedicine.

## RESULTS AND DISCUSSION

### Preparation and mechanism of liquid metal molecular scissors

Living bodies are composed of substances with different functions, which are composed of various molecules with specific elements, structures and chemical bonds. [13] As shown in Figure 1a, the common substances in life are mainly divided into organic or inorganic molecular substances, including $H_2O$ and $CH_3OH$, etc. In addition, the molecular substances also can be divided into solid, liquid and gas such as correspondingly vitamin C, $CH_3COOH$, and $CO_2$. Most molecular substances are rich in C, H, O and other elements. [14] Figure 1b displays the eGaIn exposed to molecular substances, which can spontaneously and rapidly capture the elements with strong oxidization in molecular substances, such as oxygen. When molecular materials are clipped, it would lead to the fact that the stability of the original chemical bonds was broken, and the remaining groups can then be recombined to form new substances. [15] As schematically illustrated in Figure 1c, when liquid metals are immersed in deionized water, bubbles are generated randomly at the contact interface between liquid metals and water. The verification of liquid metal surface activity editing water molecules to produce hydrogen was achieved via two processes: (i) The metal gallium was mixed with indium to form room temperature liquid metal of eGaIn. (ii) The eGaIn reacted with $H_2O$ to generate a gas. As schematically shown in Figure 1d, eGaIn is immersed in the methanol with Ar protection, bubbles are generated randomly at the contact interface between eGaIn and methanol. The metals on the surface of LMs was transformed into oxides with blackish-gray, and the supernatant presented obvious fluorescence properties under 254 nm ultraviolet excitation. According to a schematic in Figure 1e, the molecular matters are clipped by gallium molecular scissors via oxygen capture. The surface of eGaIn directionally snatches oxygen atoms from molecular substances to break up the stability of chemical bonds. And then, the remaining groups recombine to form new functional substances. [16]

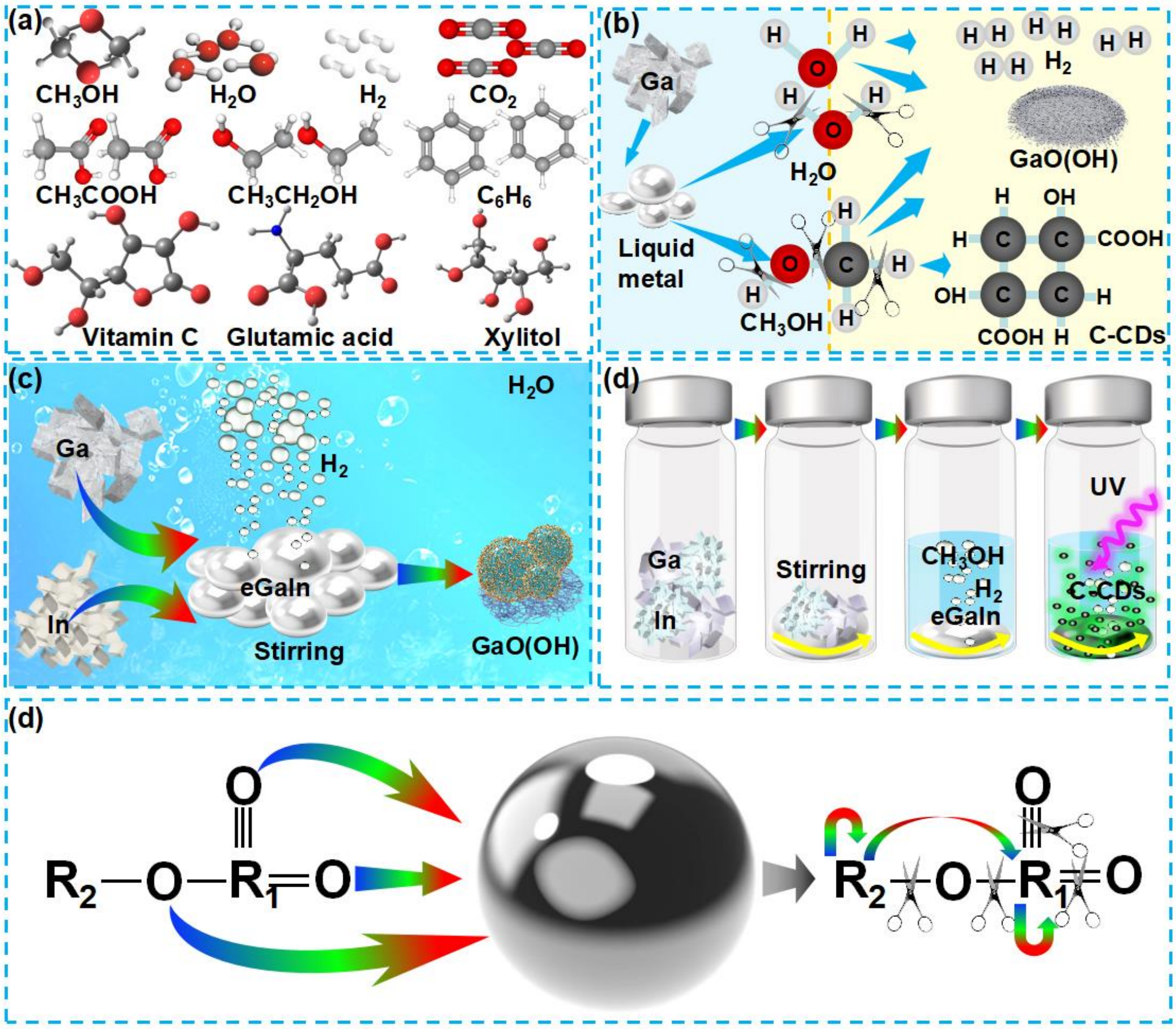

**Figure 1 The preparation process and mechanism of gallium molecular scissors via oxygen capture mechanism.** (a) The structures, element types, and chemical bonds of common molecular substances. (b) The mechanism and process of gallium molecular scissors via oxygen capture and recombining residual groups. (c) The process and mechanism of water molecular clipping with gallium molecular scissors. (d) The process and mechanism of methanol molecular clipping with gallium molecular scissors. (e) The mechanism model of gallium molecular scissors via oxygen capture.

## Inorganic molecules editing through liquid metal scissors

To disclose the editing capability of gallium-based liquid metals on $H_2O$ molecule at room temperature, eGaIn particles were immersed in deionized water and protected by Ar. Figure 2a displays that the surface of LMs was transformed into oxides, which then were stripped from the surface of liquid metal and make the deionized water shown cloudy and blackish-gray. Furthermore, the bubbles were generated randomly on the surface of water. This suggests that the eGaIn triggers the breakdown of water molecules at room temperature. As shown in Figure 2b, the gas composition in the reaction bottle is detected by a gas chromatograph, resulting in that with the increase of stirring time, the intensity of hydrogen chromatographic peak gradually increases, while the oxygen ($O_2$) chromatographic peak decreases until disappears, and the nitrogen ($N_2$) chromatographic peak changes little. This indicates that eGaIn produces hydrogen after being exposed to water ambient, absorbs oxygen, but has little effect on nitrogen. As schematically illustrated in Figure 2c, the hydrogen content in the reaction bottles is compared and calculated, indicating that the hydrogen content has a strong positive correlation with the stirring time. These proved that eGaIn liquid metal displays obvious clipping effect on water molecules at room temperature, and the remaining hydrogen atoms combine to form $H_2$. [17] As shown in Figure 2 (d-e), the solid substances of eGaIn after reacting with water molecules is determined as GaO(OH) by IR and Raman spectrum analysis, [18-19] and most of GaO(OH) is converted into $Ga_2O_3$ after annealing at 600 °C for 2.0 h, as depicted in Eqs. (1) - (2). This suggests that the eGaIn has excellent cutting capability on H-O bond in $H_2O$ molecule.

$$Ga_{(l)}+H_2O_{(l)}=GaO(OH)_{(s)}+H_{2(g)} \quad (1)$$

$$2GaO(OH)_{(s)}=Ga_2O_{3(s)}+H_2O_{(g)} \quad (2)$$

The XPS spectra of the samples were further collected to study the chemical states of each element in reaction products. [20] As shown in Figure 2f, the spectrum of Ga 2p is fitted into four sub-peaks, Ga2$p_{3/2}$ at 1115.6 eV and 2$p_{1/2}$ at 1142.44 eV, $Ga^{3+}$ 2$p_{3/2}$ at 1118.08 eV and 2$p_{1/2}$ at 1142.44 eV. Figure 2g displays that the In 3d spectrum is fitted into four sub-peaks, In 3$d_{5/2}$ at 442.73 eV and 3$d_{3/2}$ at 450.26 eV, $In^{3+}$ 3$d_{5/2}$ at 443.77 eV and 3$d_{3/2}$ at 451.22 eV, respectively. According to Figure 2h, the O 1s spectrum is fitted into two sub-peaks, Ga-O at 530.78 eV and Ga-OH at 531.71 eV. As shown in Figure 2i, the products of the eGaIn liquid metal reacted with water were annealed at 600° C for 2.0h, the proportion of Ga-O bond increased and the proportion of Ga-OH bond decreased. It was further confirmed that annealing transformed GaO(OH) into $Ga_2O_3$. [21] This indicates that gallium scissors serve well to generate hydrogen by trapping oxygen from $H_2O$, and converted gallium into functional semiconductors of GaO(OH) and β-$Ga_2O_3$. It has pretty practical values in the fields of energy, electronic information, optoelectronic devices and biomedicine.

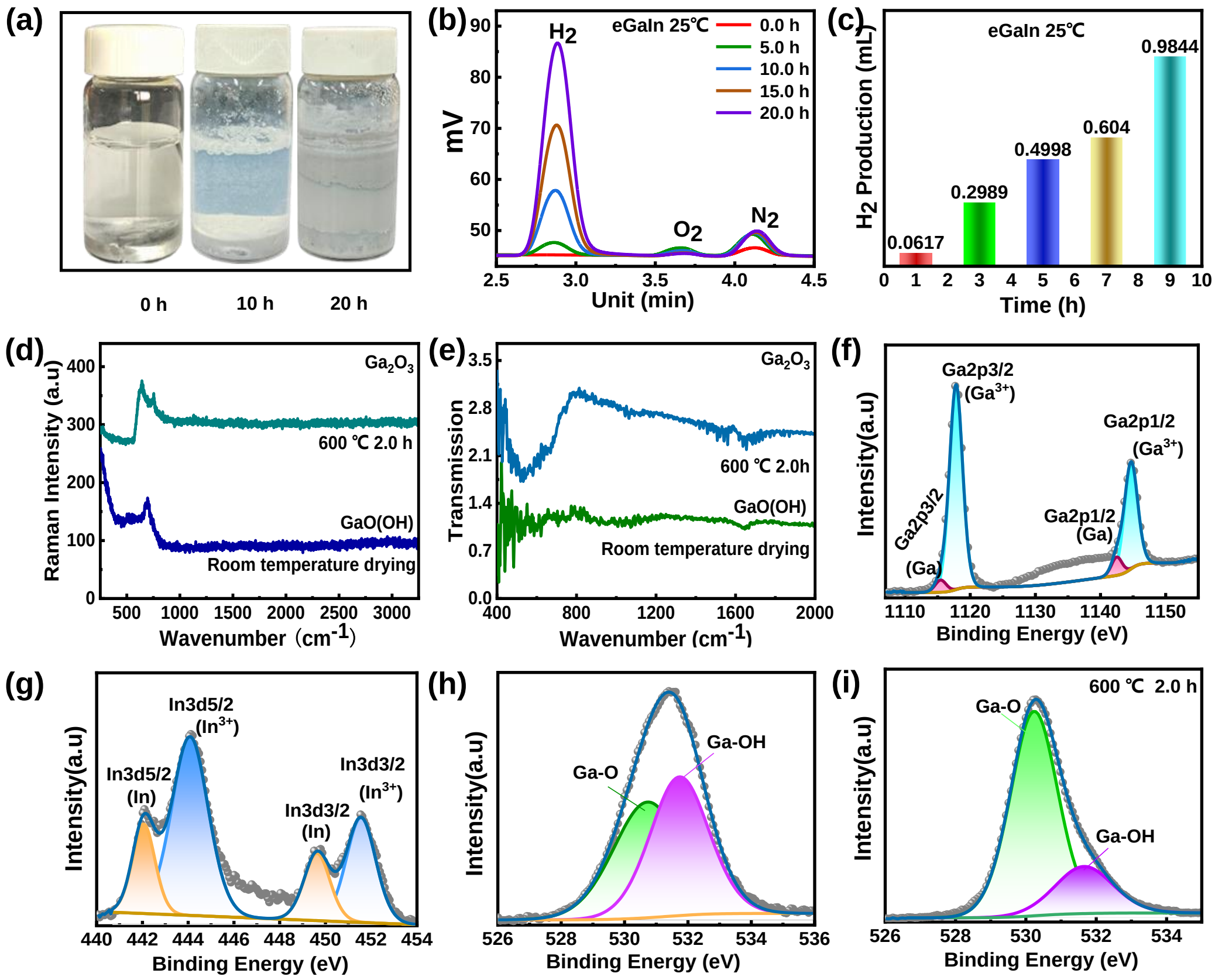

**Figure 2 The effect of gallium molecular scissors on $H_2O$ molecules.** (a) Appearance of eGaIn reacting with $H_2O$. (b) Gas chromatogram of eGaIn reacting with $H_2O$. (c) The hydrogen production of eGaIn reacting with water varying with time. (d) IR spectra of GaO(OH) and β-$Ga_2O_3$. (e) Raman spectrograph of GaO(OH) and β-$Ga_2O_3$. (f) XPS spectra and fitting result for Ga2p. (g) XPS spectra and fitting result for In3d. (h) XPS spectra and fitting result for O1s. (i) XPS spectra and fitting result for O1s of the products by eGaIn reacting with water after annealing.

As schematically illustrated in Figure 3a, liquid metals and their immersion in water all lacked crystalline lattice. The crystalline of products were determined by X-ray diffraction (XRD) analysis, and the corresponding data is depicted in Figure 3b. The results showed that GaO(OH) was mainly generated after the eGaIn liquid metal was clipped to water molecules, but GaO(OH) was gradually transformed into β-$Ga_2O_3$ after annealing at 600 °C for 2.0 h (Figure 3c). These all confirmed that Ga on the surface of eGaIn liquid metal has excellent and directional editing effect via oxygen capture for $H_2O$ molecule, and the remaining hydrogen atoms rebuild to form hydrogen. As shown in Figure 3d, the core-shell structure is an important and ubiquitous characteristic of liquid metals. Once exposed to even trace amounts of water, a thin layer of oxidation would form on the surface of liquid metals almost immediately.[22] As depicted in Figure 3e, after centrifugation and vacuum drying, the solid reaction products from eGaIn liquid metal editing water are mainly powdered, and their surfaces are covered by linear structural materials. As shown in Figure 3f, TEM observation, the supernatant of eGaIn clipping water molecules is mainly GaO(OH) nanowires with a diameter of about 10-25 nm.[23] Therefore, the gallium molecular scissors via oxygen capture editing water molecules is also an effective strategy for fabricating semiconductor nanomaterials.

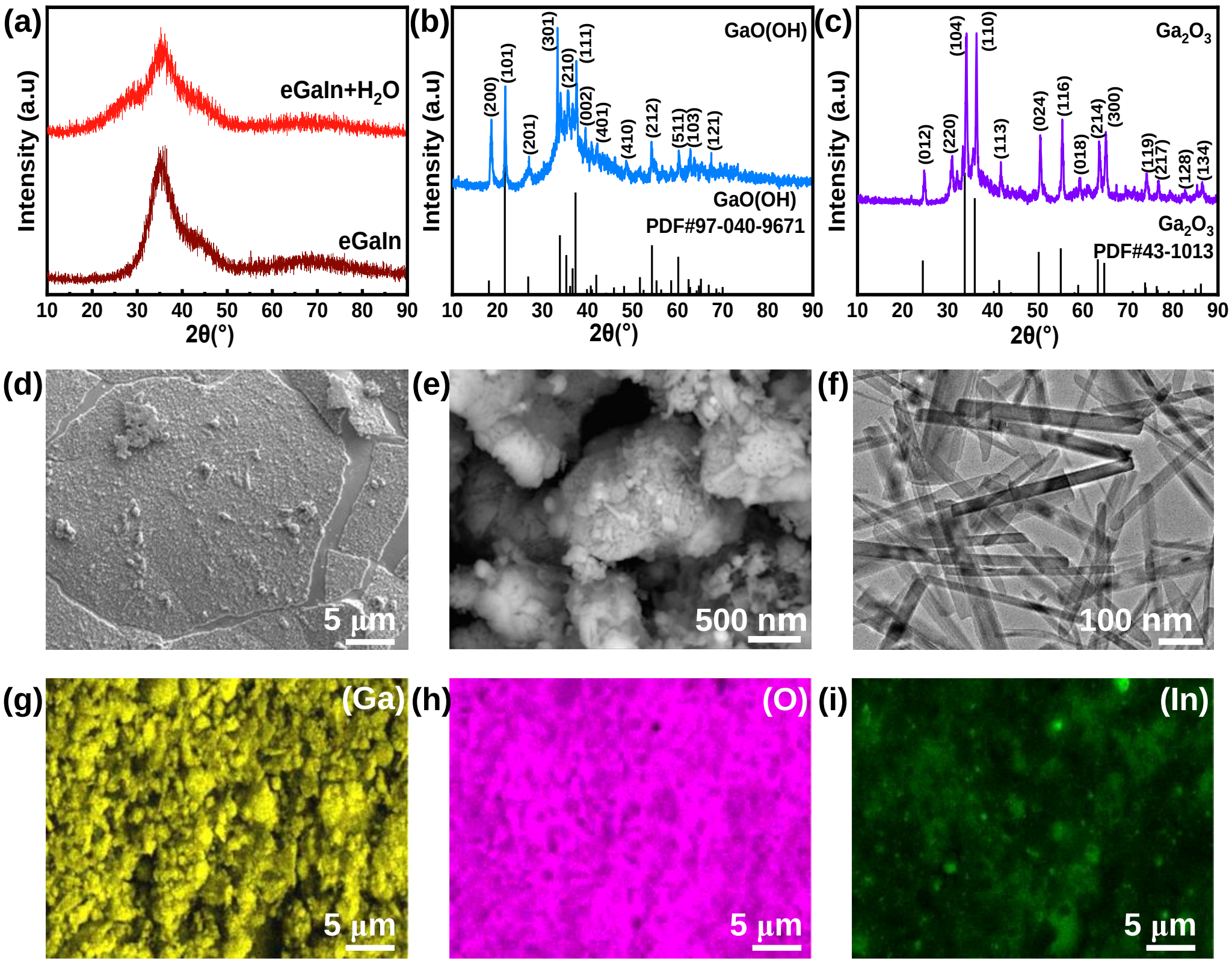

**Figure 3** **Editing effect of liquid metal scissors on water molecules at room temperature.** (a) XRD patterns of liquid metals and their immersion in water. (b) XRD patterns of the products by eGaIn reacting with water. (c) XRD patterns of the products by eGaIn liquid metal reacting with water after annealing at 600 °C for 2.0 h. (d) SEM image of eGaIn droplet exposed to the water. (e) SEM image of GaO(OH) powders. (f) TEM image of GaO(OH) nanowires. (g-i) EDS spectra of products by eGaIn liquid metal reacting with water.

## Organic molecules editing through liquid metal scissors

Methanol is a typical oxygen-containing organic molecule, and a vital chemical raw material. [24] The editing function of gallium-based liquid metal to organic molecules was further disclosed. As performed in our experiments, eGaIn was soaked in methanol and stirred continuously. As shown in Figure 4a, eGaIn gradually becomes turbid in methanol solution, accompanied by bubbles generation. Clearly, the methanol solution was clipped by gallium-based liquid metal scissors, and the remaining groups recombine to form new products and exhibit obvious fluorescence characteristics under 254nm ultraviolet excitation. This suggests that the eGaIn has a significant editing effect on methanol molecule, which allowed them to transform into new substances with fluorescence properties at room temperature. [25] As shown in Figure 4b, the gas composition in reaction bottles were analyzed by a gas chromatograph. The results indicated that the intensity of hydrogen chromatographic peak gradually increases with the stirring time. [26] In addition, ultrasonic treatment was more favorable for eGaIn to clip methanol and increase hydrogen content, while the intensity of oxygen chromatographic peak gradually decreases until disappears, and the intensity of nitrogen chromatographic peak changes little. This suggests that exposure of the liquid metal to oxygen-containing organic molecules breaks the stability of their chemical bonds through oxygen capture, and free hydrogen atoms inter-combined to form hydrogen gas. As shown in Figure 4c, the hydrogen content in the reaction bottles is compared and calculated, resulting in that the hydrogen content gradually increases with the stirring time. This proved that eGaIn has a significant cutting effect on methanol molecules by oxygen capture at room temperature. As shown in Figure 4d, the phase structures for the reaction products of eGaIn reacting with methanol were analyzed, resulting in that the products mainly display the diffraction bulge of the material structures of carbon and GaO(OH), which confirmed that the crystallinity of the products were poor. As schematically illustrated in Figure 4e, IR analysis, the reaction products of eGaIn and methanol mainly included GaO(OH), C-O-C, C-C, C-H, $COO^-$, C-O, and $OH^-$ groups, [27] and we infer that the new products with fluorescence characteristics are carbon quantum dots (C-CDs). [28] As shown in Figure 4f, Raman spectrum analysis, it is further confirmed that the products of eGaIn liquid metal reacting with methanol mainly include GaO(OH) and carbon materials. [29] This confirmed that eGaIn liquid metal can lead to the formation of carbon quantum dots with fluorescence behaviors after editing methanol molecule. The XPS spectra of the samples were further collected to study the chemical states of each element in reaction products, and the corresponding data is shown in Figure 4(g-i). As indicated in Figure 4g, the spectrum of Ga 2p is fitted into two sub-peaks, $Ga^{3+}$ $2p_{3/2}$ at 1117.74 eV and $2p_{1/2}$ at 1144.56 eV. Figure 4h displays that the O 1s spectrum is fitted into two sub-peaks, Ga-O at 530.69 eV, -OH at 532.14 eV, and $OH^-$ at 532.64 eV. As shown in Figure 4i, C 1s spectrum is fitted into three sub-peaks, C-C at 284.8 eV, C-OH at 286.42 eV, and $COOH^-$ at 288.78 eV. According to XPS analysis, it further confirmed that after the eGaIn liquid metal reacting with methanol, the oxygen-containing groups and the gallium on the surface of eGaIn formed gallium oxides, a variety of new substances can be formed through random combinations of remaining groups, including hydrogen, carbon quantum dots, etc. The cutting process can be described as Eqs. (3):

$$Ga_{(l)}+CH_3OH_{(l)}=GaO(OH)_{(s)}+H_{2(g)}+C\text{-}CDs(s) \quad (3)$$

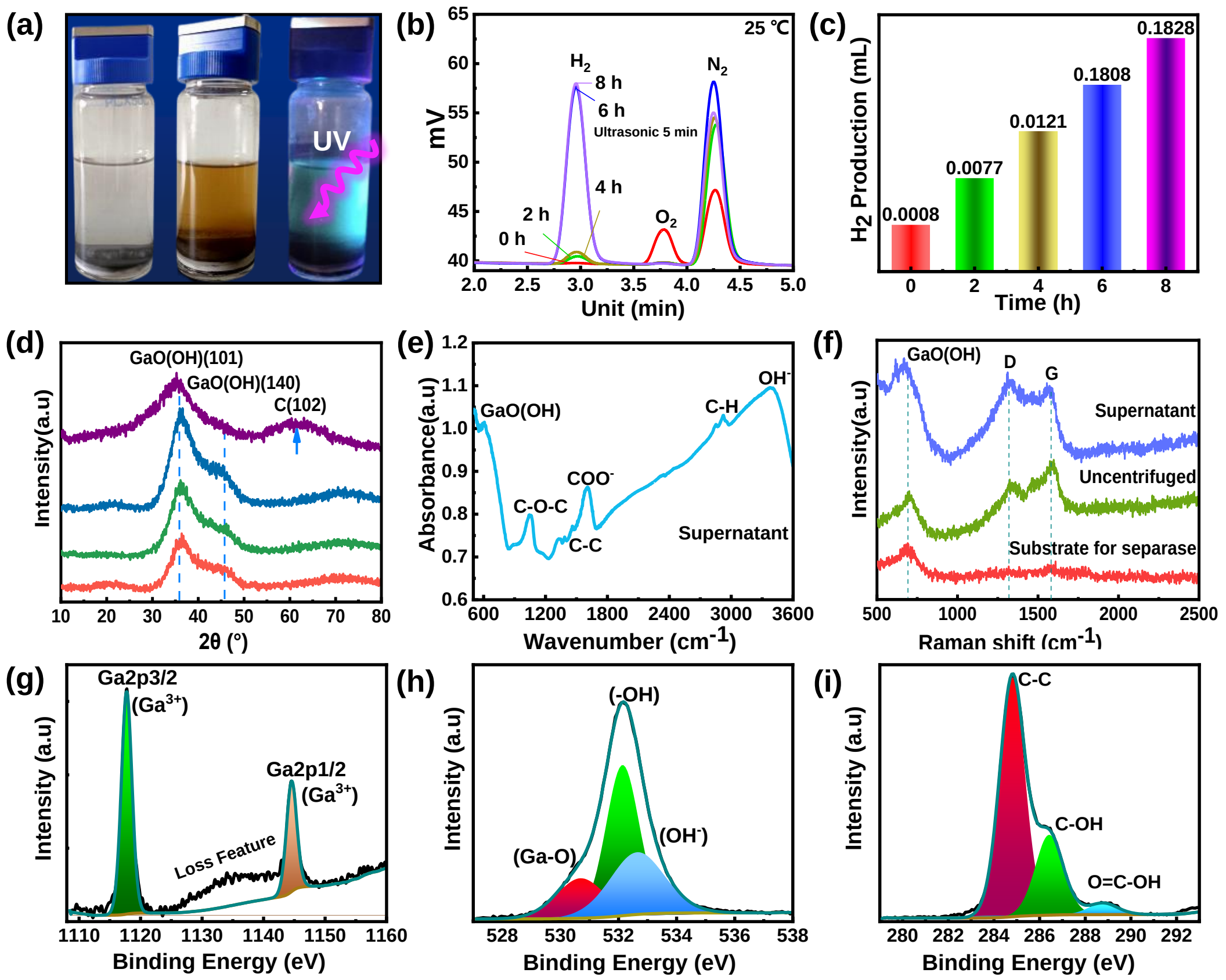

**Figure 4 Editing effect of gallium molecular scissors on methanol molecules.** (a) Appearance of eGaIn reacting with methanol. (b) Gas chromatogram of eGaIn reacting with methanol. (c) Hydrogen production of eGaIn reacting with methanol with time. (d) XRD patterns of products by eGaIn reacting with methanol. (e) IR spectra of products by eGaIn reacting with methanol. (f) Raman spectra of products by eGaIn reacting with methanol. (g) XPS spectra and fitting result for Ga2p. (h) XPS spectra and fitting result for O1s. (i) XPS spectra and fitting result for C1s.

The methanol solution presented obvious fluorescence characteristics under UV excitation after eGaIn editing. Figure 5a showed the photoluminescence (PL) spectra for all samples measured in the region around 500 nm. The intensities and positions of the PL peaks shifted with changes in reaction time. The PL spectra was fitted using Gaussian model, revealing multiple light emission peaks corresponding to blue, green, yellow, and red wavelengths, respectively (Figure 5b). Figure 5c illustrated variations in the fluorescence decay profiles in the same strategy, the PL decay spectra can be fitted by a bi-exponential function providing in real time the ($\tau_1$) and ($\tau_2$) components, [30] and the average recombination lifetime. Figure 5c indicated that the maximum fluorescence lifetime of methanol solution after eGaIn editing is 2.51 ns. The position, area, and FWHM (Full-width half-maximum) of the peaks varied with the reaction time, which was consistent with the shifts in the relevant CIE coordinates. Figure 5d depicted the CIE chromaticity diagrams of the three samples, and the shift in luminescence color with reaction time. The CIE coordinates were correspondingly CIEx,y = 0.41, 0.53, CIEx,y = 0.30, 0.61, and CIEx,y = 0.31, 0.61, respectively. These suggested that the types and properties of the methanol molecular editing products as tested have a strong dependence on time and methods. As shown in Figure 5e, after vacuum drying, a core-shell structure is formed between the reaction products, which is tightly connected and not easy to fall off. Further, the contents of Ga, O, C, and In elements in the core-shell structure are shown in Figure 5f. The shells and their texture are mainly rich in C (72.7 wt.%), with little amounts of Ga (16.9 wt.%) and O (10.4 wt.%). But the core and their texture are mainly rich in Ga (65.9 wt.%) and O (26.9 wt.%), with little amounts of C (7.2wt.%). This suggests that the cores are mainly composed of GaO(OH), and the shells are formed by the deposition of carbon quantum dots during the drying process. As shown in Figure 5g, the core-shell structure formed by Ga(OH) and carbon materials is about 10-50 nm in size. Figure 5g also displays the EDS mapping acquired on the core-shell structure particles, resulting in that the distribution of the Ga, O, and C elements on the surface of core-shell structure particles was relatively broad. The EDS spectrum also showed enrichment of C, O elements with significant correlation in shell region, and the Ga, O elements overlapped with the enriched region in core. This indicated that after eGaIn combing with methanol molecule, the thin layers stacked with Ga(OH) and carbon quantum dots were formed instantaneously and spontaneously at their contact interface, meanwhile hydrogen gas is released. The composite shells of Ga(OH) and carbon quantum dots covered on the surface of eGaIn, which prevented the liquid metal particles from binding to each other during agitation. In addition, combined with all the studies and analyses, the outer layer of the carbon quantum dot shell also adsorbed -H, -OH, -COOH and other groups.

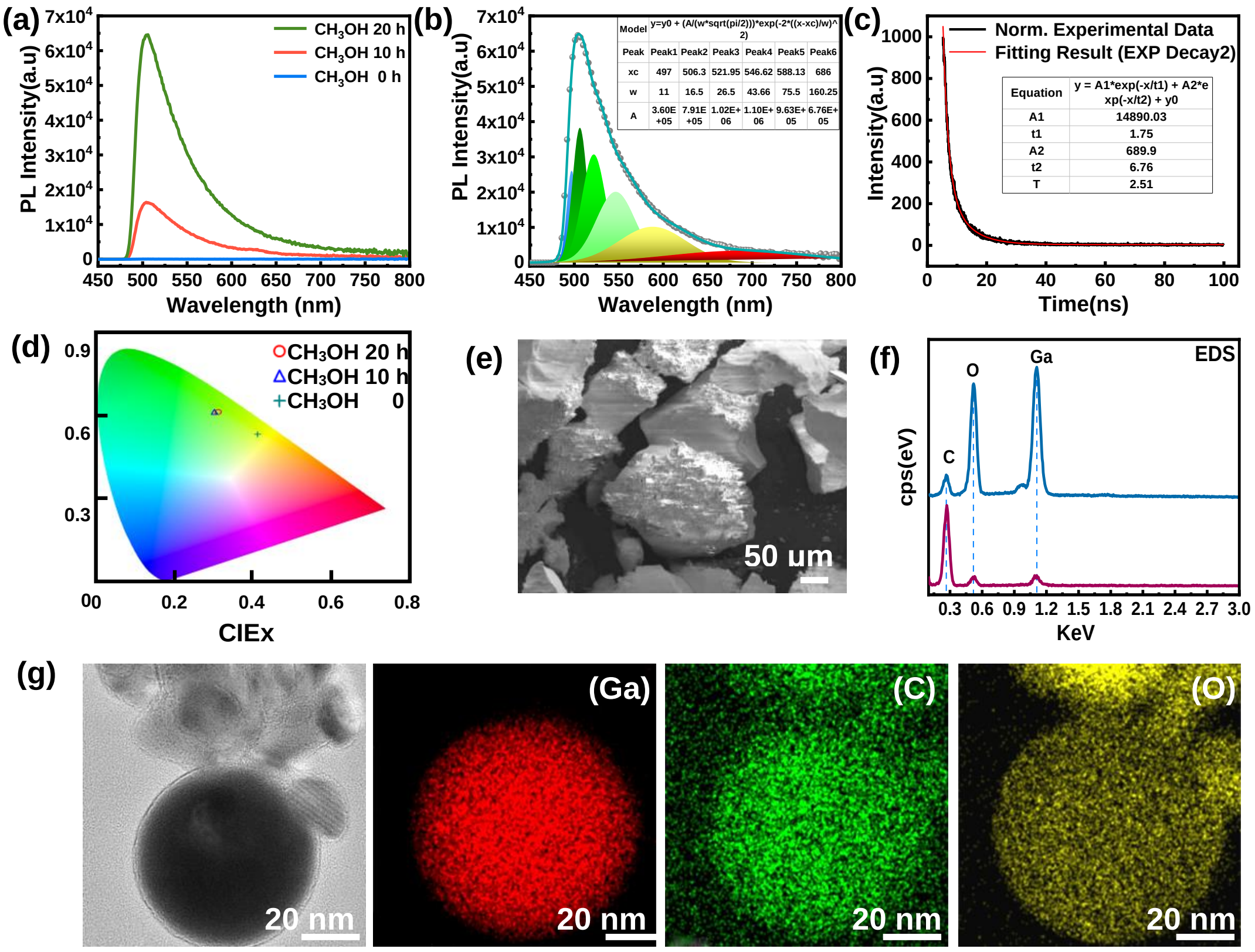

**Figure 5 Luminescence properties and micromorphology of eGaIn reacting with methanol.** (a) PL spectra of eGaIn reacting with methanol. (b) Multi-peak fitting with Gaussian model of eGaIn

reacting with methanol fitted using a biexponential model. (c) TPL decays fitted using a bi-exponential model. (d) CIE coordinates of eGaIn reacting with methanol. (e) SEM image for the core-shell structure of reaction products. (f) EDS spectra for the core-shell structure of reaction products. (g) TEM and EDS spectra for the core-shell structure of reaction products.

## CONCLUSION

In summary, the phase structures and surface chemical activity of room temperature liquid metals are essential for achieving unique molecular scissors on demand. A basic way was discovered and demonstrated to realize the gallium molecular scissors via oxygen capture. Except for the tested eGaIn liquid materials by mixing of Ga with In particles under Ar protection, more scissors can also be developed in the coming time. At this stage, the transition of the liquid phase structure significantly increased the surface chemical activity of Ga, leading to its spontaneous and rapid capture of oxygen atoms in oxygen-containing molecular matters. As a typical experimental demonstration, the eGaIn spontaneously and rapidly reacted with water to form GaO(OH), along with hydrogen production. This outputs quite useful byproduct GaO(OH), which is in fact a semiconductor material with convenient structure and performance regulation. We also found that the eGaIn could react with methanol to generate GaO(OH), $H_2$, carbon quantum dots, and other related products. It should be pointed out that, gallium molecular scissors via oxygen capture is a spontaneous process, which avoids the traditional harsh conditions such as high temperature, high vacuum and catalyst, breaking through the limitations of performance and application singularization facing many molecular substances. Clearly, the abundance of organic and inorganic molecules, and the fluidity and activity of gallium-based liquid metals guarantee tremendous opportunities for tackling the tough challenges in diverse areas from energy, environment, to advanced materials etc. This work established a general, tunable, and scalable liquid metal scissors for cutting and editing molecules via a rather easygoing way. The scissors are also expected to be a powerful tool for innovating future chemical engineering, life science, energy and environment, and biomedicine where various molecules are playing indispensable roles everywhere.

## MATERIALS AND METHODS

### Syntheses and processing of materials

The LMs (eutectic gallium and indium, eGaIn) were made by high purity gallium (Ga) and indium (In) metals (with purity of 99.9%) as raw materials. The mass ratios of Ga, In in the eGaIn were 75.5% and 24.5%, respectively. The Ga and In metals were determined by an electronic balance (METTLER TOLEDO 1.0 $\times 10^{-5}$ g). Then the pre-weighed Ga and In metals were mixed and stirred to prepare the liquid metals at room temperature for 30 min in argon (Ar) protection.

### Processing of molecular editing

Gallium molecular scissors via oxygen capture triggers the molecular clipping at room temperature which was synthesized by the following steps: Firstly, the 50 ml glass reaction bottles were filled with 30 ml $H_2O$ or $CH_3OH$, respectively. Secondly, the air in the reaction bottles was removed using a gas controller, and the argon (Ar) was filled as a protective gas to isolate the external air into reaction bottles. Then the eGaIn (10g) is injected into $H_2O$ or $CH_3OH$ from the seal of reaction bottles. Finally, The eGaIn is constantly exposed to $H_2O$ or $CH_3OH$ to produce hydrogen ($H_2$) and any solid materials.

**Characterization methods**

The X-ray diffraction (XRD) patterns were acquired on a SmartLabSE X-ray diffractometer (CuK$\alpha$1, k = 1.54056 Å, 40 kV, 50 mA), in the scanning range of $2\theta$ = 5° to 90°. X-ray photoelectron spectra (XPS) were acquired using ESCALAB250Xi X-ray photoelectron spectrometer (USA), wherein the binding energy of the C 1s peak at 284.8 eV was used as the internal reference. Scanning electron microscopy (SEM) and energy dispersive X-ray spectroscopy (EDS) were conducted on a TESCAN AMBER and Nova NanoSEM 450 scanning-electron microscope at 20 keV voltage. The infrared spectroscopy patterns were acquired on a FTIR-2000 Fourier transform infrared (FTIR) spectroscopy, and the scanning range is 400 $cm^{-1}$ to 4400 $cm^{-1}$. The Raman spectroscopy patterns were acquired on a LabRam HR Evolution high-resolution confocal laser micro-Raman spectrometer, and the scanning range is 200 $cm^{-1}$ to 3250 $cm^{-1}$. The gas in the reaction bottle was detected and analyzed by a GC9790II gas chromatography. Photoluminescence (PL) spectra were obtained using a FLS 1000 fluorescence spectrometer (Edinburgh Instruments, UK), with excitation wavelength of 254-365 nm, scanning speed of 1 nm $s^{-1}$, and the widths of both the excitation slit and emission slit are 2.0 nm.

# Acknowledgement

This work was funded by the National Key Research and Development Program of China (2022YFB3803600), National Natural Science Foundation of China Projects (51890893), the Key Research and Development Program of Yunnan Province (202302AF080002), Yunnan Yunling Scholars Project, Young and Middle-aged Academic and Technical Leaders Reserve Talent Project in Yunnan Province (202005AC160015). Yunnan Industrial Technology Innovation Reserve Talent Project (No. 202105AD160056). Yunnan Basic Applied Research Project (No. 202101AT070013, 2019-1-C-25318000002171). Authors thank the Electron Microscopy Center, the Advanced Computing Center, the Advanced Analysis and Measurement Center of Yunnan University for the sample testing and computations service.

# Author contributions:

Conceptualization: L. Duan, J. Liu, Q. Liu. Methodology: L. Duan, T. Zhou. Investigation: L. Duan, T. Zhou, H. Yang, W. Mu. Project administration and supervision: Q. Liu, J. Liu. Writing – original draft: L. Duan, T. Zhou. Writing – review & editing: L. Duan, J. Liu, Q. Liu. All authors have discussed and given approval to the final version of the manuscript.

# References

1. Segev, Y., Pitzer, M., Karpov, M., et al. (2019). Collisions between cold molecules in a superconducting magnetic trap. *Nat.* **572**, 189-193.

2. Uchida, J., Yoshio, M., Kato, T. (2021). Self-healing and shape memory functions exhibited by supramolecular liquid-crystalline networks formed by combination of hydrogen bonding interactions and coordination bonding. *Chem. Sci.* **12**, 6091-6098.
3. Sun, H., Kabb, C. P., Dai, Y. Q., et al. (2017). Macromolecular metamorphosis via stimulus-induced transformations of polymer architecture. *Nat. Chem.* **9**, 817-823.
4. Zhang, Z. H., Andreassen, B. J., August, D. P., et al. (2022). Molecular weaving. *Nat. Mater.* **21**, 275-283.
5. Chou, C. W., Kurz, C., Hume, D. B., et al. (2017). Preparation and coherent manipulation of pure quantum states of a single molecular ion. *Nat.* **545**, 203-207.
6. Song, Q., Cheng, Z. H., Kariuki, M., et al. (2021). Molecular self-assembly and supramolecular chemistry of cyclic peptides. *Chem. Rev.* **121**, 13936-13995.
7. Duan, L. F., Zhou, T., Zhang, Y. M., et al. (2023). Surface optics and color effects of liquid metal materials. *Adv. Mater.* **35**, 2210515.
8. Duan, L. F., Zhang, Y. M., Zhao, J. H., et al. (2022). New strategy and excellent fluorescence property of unique core-shell structure based on liquid metals/metal halides. *Small* **22**, 2204056.
9. Zavabeti, A., Ou J. Z., Carey, B. J., et al. (2017). A liquid metal reaction environment for the room-temperature synthesis of atomically thin metal oxides. *Sci.* **358**, 6361.
10. Yang, N. L., Gong, F., Zhou, Y. K., et al. (2022). Liquid metals: Preparation, surface engineering, and biomedical applications. *Coord. Chem. Rev.* **471**, 214731.
11. Shirhatti, P. R., Rahinov, I., Golibrzuch, K., et al. (2018). Observation of the adsorption and desorption of vibrationally excited molecules on a metal surface. *Nat. Chem.* **10**, 592-598 (2018).
12. Fu, J. H., Liu, T, Y., Cui, Y. T., et al. (2021). Interfacial engineering of room temperature liquid metals. *Adv. Mater. Interfaces* **8**, 2001936.
13. Chang, R., Zhao, L. Y., Xing, R. R., et al. (2023). Functional chromopeptide nanoarchitectonics: Molecular design, self-assembly and biological applications. *Chem. Soc. Rev.* **52**, 2688-2712.
14. Cuenca, A. B., Shishido, R., Ito, H., et al. (2017). Transition-metal-free B-B and B-interelement reactions with organic molecules. *Chem. Soc. Rev.* **46**, 415-430.
15. McNally, A. (2015). Molecular editing of carbohydrates. *Nat. Chem.* **7**, 539-541.

16. Diercks, C. S., Dik, D. A., Schultz, P. G. (2021). Adding new chemistries to the central dogma of molecular biology. *Chem.* **11**, 2883-2895.

17. Zhu, X. D., Song, Z. M., Wang, Z. Y, et al. (2020). Selective formation of interfacial bonding enables superior hydrogen production in organic–inorganic hybrid cocatalyzed photocatalysts. *Appl. Catal., B.* **274**, 119010.

18. Zhao, Y. Y., Frost, R. L. (2020). Raman spectroscopy and characterisation of α-gallium oxyhydroxide and β-gallium oxide nanorods. *J. Raman Spectrosc.* **39**, 1494-1501.

19. Feneberg, M., Bläsing, J., Sekiyama, T., et al. (2019). Anisotropic phonon properties and effective electron mass in α-$Ga_2O_3$. *Appl. Phys. Lett.* **114**, 142102.

20. Duan, L. F., Zhang, Y. M., Zhao, J. H., et al. (2022). Unique surface fluorescence induced from the core-shell structure of gallium-based liquid metals prepared by thermal oxidation processing. *ACS Appl. Mater. Interfaces* **34**, 39654-39664.

21. Chao, Y. W., Liu, B. G., Zhang, L. B. (2020). Dielectric properties, structure and morphology during synthesis of β-$Ga_2O_3$ by microwave calcination of GaOOH. *Ceram. Int.* **46**, 24923-24929.

22. Ren, L., Cheng, N. Y., Man, X. K., et al. (2021). General programmable growth of hybrid core-shell nanostructures with liquid metal nanodroplets. *Adv. Mater.* **33**, 2008024.

23. Liu, Q., Zou, R., Bando, Y., et al. (2015). Nanowires sheathed inside nanotubes: Manipulation, properties and applications. *Prog. Mater. Sci.* **70**, 01-49.

24. Zhang, L. L., Fu, T. J., Ren, K., et al. (2023). Finely regulating methanol concentration to control the alkylation depth in methanol aromatization for optimizing product distribution. *Appl. Catal., B.* **321**, 122047.

25. Deng, Y. M., Li, S., Appels, L., et al. (2022). Producing hydrogen by catalytic steam reforming of methanol using non-noble metal catalysts. *J. Environ. Manage.* **321**, 116019.

26. Zhang, Y. M., Zhao, J. H., Wang, H., et al. (2022). J. Single-atom Cu anchored catalysts for photocatalytic renewable $H_2$ production with a quantum efficiency of 56%. *Nat. Commun.* **13**, 58.

27. Zhou, G., Mikinka, E., Golding, J., et al. (2020). Investigation of thermal degradation and decomposition of both pristine and damaged carbon/epoxy samples with thermal history. *Composites, Part B.* **201**, 108382.

28. Zhao, B., Ma, H. Y., Zheng, M. Y., et al. (2022). Narrow-bandwidth emissive carbon dots: A rising star in the fluorescent material family. *Carbon Energy* **4**, 88-114.

29. Panyathip, R., Sucharitakul, S., Phaduangdhitidhada, S., et al. (2021). Surface enhanced Raman scattering in graphene quantum dots grown via electrochemical process. *Mol.* **26**, 5484.

30. Zhao, R. M., Xie, L., Zhuang, R. S., et al. (2021). Interfacial defect passivation and charge carrier management for efficient perovskite solar cells via a highly crystalline small molecule. *ACS Energy Lett.* **12**, 4209-4219.